# Characterization of Phosphorite Bearing Uraniferous Anomalies of Bijawar region, Madhya Pradesh, India.

Pragya Pandit[1], Shailendra Kumar[1], Pargin Kumar[2,] Manoj Mohapatra[3]

[1]Atomic Minerals Directorate for Exploration and Research, New Delhi-110 066, India.
[2]National Institute of Technology, Jalandhar- 144011, India
[3] Radiochemistry Division, Bhabha Atomic Research Center, Mumbai- 400085, India

## Abstract

The uranium containing phosphatic shale sub surface samples collected from Bijawar region Madhya Pradesh (M.P.), India as a part of the uranium exploration programme of the Atomic Minerals Directorate for Exploration and Research (AMD), Department of Atomic Energy (DAE) were characterized by a variety of molecular spectroscopic techniques such as photoluminiscence (PL), time resolved photoluminiscence spectroscopy (TRPL), X-ray absorption near edge spectroscopy (XANES), Raman spectroscopy and structural techniques such as X-ray diffraction (XRD) to identify the oxidation state, physical and chemical form of uranium in this region. Oxidation state analysis by fluorescence spectroscopy revealed majority of the samples in U(VI) oxidation state. The most abundant form of uranium was identified as uranate ($UO_6^{6-}$) ion, as a substituent of $Ca^{++}$ in $Ca_5(PO_4)_3F$, Flourapatite (FAP). Two uranium species in U(VI) and U(IV) state; $(UO_2)^{2+}$ adsorbed on silica and uranium oxide species ($UO_2$) were also identified. The study provided baseline information on the speciation of uranium in Bijawar region.

## Keywords

# Introduction

Proterozoic Bijawar Basin exposed along the southeastern margin of the Bundelkhand massif is one of the main prospecting regions for uranium exploration of AMD [1]. The exploration programme in Bijawar region was suspended in 2016. The site contains high level of phosphate. Mobility of uranium in the environment is dependent on its speciation and physicochemical properties of the host environment which is affected by various geological, geochemical and hydrogeological aspects. Therefore, the speciation studies have profound implication in understanding the uranium retention processes. Selective spectroscopic, methods for the speciation of uranium is provided in Figure below.

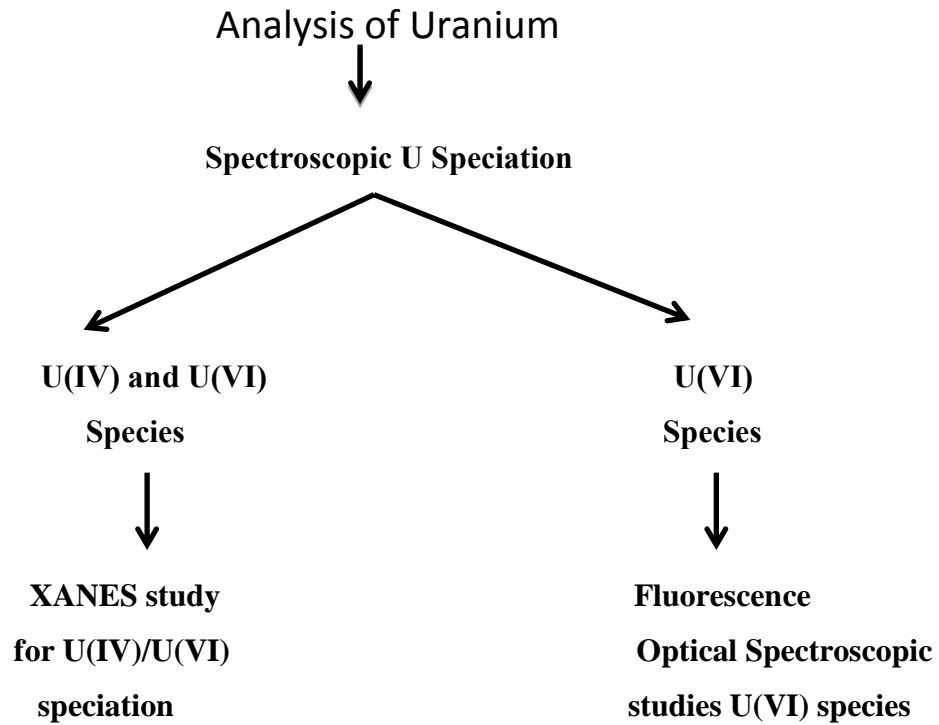

So far, there has been a lack of qualitative information on the redistribution of different oxidation states of uranium in the Bijawar region. The major analytic challenges in the speciation and spectroscopic characterization of uranium species in Bijawar region are their low concentration and interference due to the heterogeneous elements in the complex rock matrix.

Previous studies conducted have focussed on the investigation of structural and major and trace element composition in the Bijawar and Paleoproterozoic phosphorites of the Sonrai basin by XRD, scanning electron microscopy (SEM) and inductively coupled plasma emission spectroscopy (ICP-AES), however no studies so far has been available on the oxidation state and speciation of uranium in this region [2].

Fluorescence spectroscopy is a widespread tool for characterization of U(VI) species where uranium is present either as uranyl $(UO_2)^{2+}$ or uranate $(UO_6)^{6-}$ ion speciation due its high sensitivity, specificity, simplicity, ease of operation and quick response. Large amount of the literature has been available on the use of fluorescence spectroscopy for the speciation of uranium in abandoned uranium contaminated sites [3].

In the present investigations, a comprehensive photoluminescence (PL) study along with lifetime decay by Time resolved photoluminescence (TRPL) has been carried out on the samples from Bijawar region to determine uranium oxidation states and the uranium moiety associated with these oxidation states. Time resolved fluorescence spectroscopy provides a powerful tool for identifying multiple species present in a system due to their different decay time span. An attempt has been made in the present study to analyze the oxidation state of uranium in the grab and subsurface samples using fluorescence spectroscopy. Our findings substantiate the suitability of TRPL in identifying the chemical valence state of uranium in low and medium concentration. The PL results are further corroborated with XRD and Raman spectroscopy.

**GEOLOGY AND DESCRIPTION OF THE BIJAWAR REGION**

The study area constituted Proterozoic Bijawar basin exposed along the southeastern margin of the Bundelkhand massif. Bijawar Group in M.P. covers part of the Chattarpur and Sagar districts between latitudes (24º 19´ 0"N and 24º 23´ 0"N) and longitudes (79º 9´0" E and 79°14'0" E). The Paleoproterozoic metasedimentary rocks of Bijawar Group resting upon the Archean Bundelkhad Basement Complex are overlain by the rocks of Vindhyan Supergroup (Jha et al. 2012). The normal annual rainfall received in the region is 1068.3mm. The period from March to middle June constitutes the summer season, south west monsoon runs from middle of June to September followed by cold season from December to February.

The sedimentary Proterozoic Bijawar basin exposed along the south eastern margin of the Bundelkhand massif of Peninsular India constitute favourable target areas for hosting fracture-controlled hydrothermal-type and unconformity-related uranium deposits. The Bijawar Group of rocks have been categorised petromineralogcally into phosphatic ferruginous breccias, phosphatic breccias, cherty ferruginous phosphorites and ferruginous phosphatic shale. Uranium mineralization in Bijawar basin is hosted by phosphatic ferrouginous shale breccia and the average concentration of uranium varies between approximately 150 ppm to 0.1%. Significant uraniferous anomalies were located in the field season (F.S.) 2014-15 along the unconformity between Lower Proterozoic Karri Ferruginous Formation (KFF) of upper Bijawar Group in the north unconformably overlain by Middle Proterozoic Semri Group of rocks in the south. The grab and subsurface samples for study were selected from the Bijawar region. The mineralistion was hosted by ferruginous shale breccia with angular and sub rounded fragments of quartz in KFF. The subsurface samples analysed upto 0.159% $U_3O_8$ maximum value with leachable uranium and negligible thorium.

## Experimental

Four shallow subsurface SH-9, SH-15, SH-17, SH-22 (depth<2m) and two uraniferous species SH-a and SH-b were collected during the F.S. 2014-2015 following the reconnoissance survey. The uranium concentration in the background level of the sampling area was typically 5-7 ppm. The samples were selected based on the >3 BG. The samples were primarily unconsolidated. The important baseline information on the Bijawar Vindhyan Investigation was taken from the petromineralogical reports of F.S. 2013-2015. Table 1 summarises the geological coordinates, physical and chemical characteristics of the sample. PH of the samples was in the range from 6.86 –7.24 indicating neutral nature of the soil matrix due to the association of $H^+$ and $OH^-$ ions with the ion complexes. The total organic content of the sample was in the range from 2.4% to 10%.

The representative subsamples of 100 gm quantity from each of the lots were powdered (mesh size -60) and homogenized by coning and quartering. The soil matrix predominantly consisted of apatite (density 3.1-3.5 gm/cm$^3$), calcite (density 2.71 gm/cm$^3$), FAP (density 3.1- 3.3 gm/cm$^3$), quartz(density 3.1- 3.3 gm/cm$^3$) and clay(density 3.1- 3.3 gm/cm$^3$). Most of the uranium minerals have a density > 8.34 gm/cm$^3$. Visual inspection of the samples under UV light revealed no visible light emission due to any particulates in SH-9, SH-15, SH-17, SH-22 samples. SH-a showed green emission under UV excitation. Hence there is probability of complexation of uranium ions with the cationic and anionic ligands or being adsorbed on silica. Density and particle size fractionation was therefore carried out using bromoform ($\rho$=2.89 g/cm$^3$) and methylene iodide ($\rho$=3.35 g/cm$^3$) separation to selectively concentrate uranium rich species. The bromolight portion of the samples consisting of quartz, mica and clay minerals were discarded and bromoheavy portion containing uranium species was selected for analysis. Phase purity and crystal structure of the samples were identified by X ray diffractometer using Cu K$\alpha_1$ radiation (1.5406 Å) fitted with a Nickel filter (Rigaku). XRD data was collected in θ- 2θ mode from 20° to 70° with a step size of 0.02°. Simulation of crystal structure based on measured XRD data was carried out using refinement software. Raman spectra were obtained using a Raman spectrometer with an excitation source of 785 nm with laser power of 1 mw with spectral resolution of 1 cm$^{-1}$(Reinshaw Invia). The photoluminescence (PL& PLE) spectra were measured using a FLS980 fluorescence spectrometer (Edinburgh Instruments Ltd). For excitation, a 450W ozone free xenon arc lamp with tunable wavelength has been used as a standard light source. The measurements were performed with a step size of 0.1 nm, dwell time of 0.3 seconds and slit width of 0.5 nm. Time-resolved photoluminescence decay (TRPL) measurements in the nanosecond and microsecond decay scales were performed by nanosecond (ns) pulse diode laser ($\lambda_{exc}$ = 266 nm, pulse duration 260 ps and repetition frequency 20 kHz) and a xenon microsecond (μs) flash lamp with 10-100 Hz frequency. Decay time measurements were performed by using time correlated single photon counting technique.

**Results and discussion**

*X-ray Diffraction*

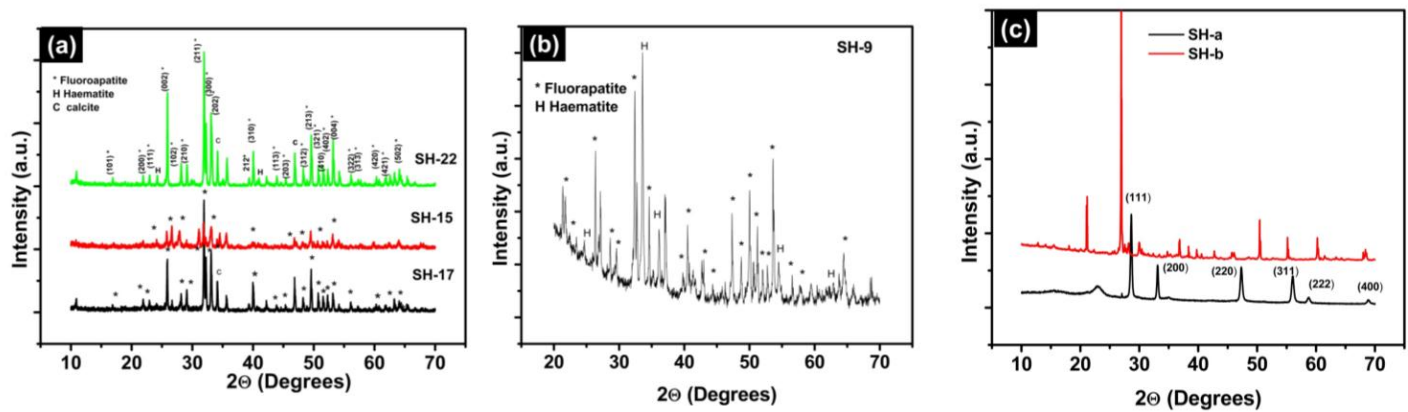

**Fig. 1** XRD curve of (a) SH-15, SH-17& SH-22, (b) SH-9 and (c) SH-a and SH-b

The chemical composition of the samples determined by inductively coupled plasma emission (ICP-AES) revealed the main constituents to be CaO (40.4%), $P_2O_5$ (33.0%), $Fe_2O_3$ (11.7 %), $SiO_2$ (20.8%) along with trace elements in ppm concentration. XRD data for four powdered samples SH-15, SH-17 and SH-22, SH-9 and uraniferous species SH-a and SH-b are illustrated in Fig. 1. XRD data for the Bijawar region has been collected by different groups. The whole rock samples of SH-15, SH-17, SH-22 and SH-9 indicated FAP to be the predominant phase in the XRD. (Fig. 1(a)). The observed 2θ reflections at 25.89°, 29.01°, 31.90°, 32.20°, 33.10°, 34.16°, 39.94°, 46.87°, 49.57° and 51.41° corresponding to the hkl planes (002), (120), (121), (112), (300), (202), (310), (222), (123) and (410) were observed in agreement with the JCPDS card no [15-0876]. Along with it guange phases corresponding to calcite, $CaCO_3$ [JCPDS # 05-0586] and haematite, $Fe_2O_3$ [JCPDS #33-0664] were also observed. The experimental XRD patterns were simulated to obtain the lattice parameters in the range of a= 9.42 Å, c = 6.75 Å, α = $90^0$, β= $90^0$ and γ=$120^0$, lattice volume; V= 525.96 Å³ and space group $P_{63/m}$ [4]. No reflections corresponding to any uranium minerals were observed. The XRD data for SH-9 revealed peaks corresponding to FAP and those at 24.23°, 33.19°, 40.89° and 54.19 °corresponding to (012), (104), (113) and (116) planes of haematite (Figure 1(b)). Figure 1(c) represents the XRD data for the species SH-a and SH-b. The X-ray spectral data for the species SH-a was found similar to that with silica ($SiO_2$) and chalcopyrite. The 2θ peaks at 20.82°, 26.62°, 30.06°, 36.86° and 50.10° corresponding to (100), (011), (103) (110) and (112) reflections of crystalline quartz was observed [JCPDS # 86-1630]. The other reflections obtained were due to the guange phase of chalcopyrite. The diffraction data from SH-b revealed spectral similarity with uranium oxide; $UO_2$ [JCPDS # 41-1422]. The structure was identified to be cubic and the lattice parameters observed were a= b=c=5.375Å [5].

*Raman spectroscopy*

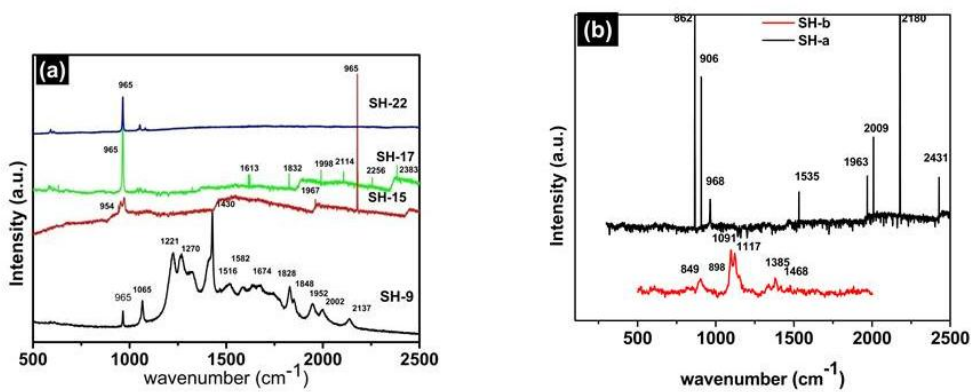

**Fig. 2** Raman curves of (a) SH-9, SH-15, SH-17 &SH-22 (b) SH-a &SH-b

Raman spectroscopy could reveal information about the bonding environment of the molecules and is a widespread tool for analyzing geological samples. Raman spectra of the samples over the spectral range from 500 to 2500 cm$^{-1}$ is shown in the Figure 2. The Raman spectrum for the samples SH-15, SH-17, SH-22 and SH-9 corresponds to a well crystallized structure. The peak for the phosphate (PO$_4$) group characteristic of the $\nu_1$ asymmetric stretching is observed in the range from 950- 970 cm$^{-1}$ in all of the samples [6]. The other bands appearing in the range of 2134 cm$^{-1}$ corresponds to organic content while those corresponding to 1673 and 1674 are those of water molecules. Raman spectra of SH-9 exhibited bands due to haematite, PO$_4$ and CO$_3^{2-}$ group and the results are discussed in our earlier report. SH-b exhibits the Raman spectra of the uraninite species similar to that observed by Stefaniak et al. [7]. The Raman spectra of SH-a in (Figure 2(b)) reveal the characteristic symmetric stretching and antisymmetric stretching band corresponding to the UO$_2^{2+}$ in the vicinity around 860 cm$^{-1}$ and at higher wavenumber at 907 cm$^{-1}$ [8]. This is similar to the vibrational properties exhibited by uranyl minerals. Raman spectra of SH-b revealed peaks at 849, 898, 1091, 1117, 1385 and 1468 cm$^{-1}$. The Raman peaks at 898 and 1117 cm$^{-1}$ may be assigned to the symmetric stretching bands of (UO$_2$)$^{2+}$ and (CO$_3$)$^{2-}$ group. Different studies conducted on various uraninite minerals has shown that under different geochemical condition alteration of natural urannite takes place resulting in uranium silicates, oxides or carbonates [9].

*Photoluminescence studies*

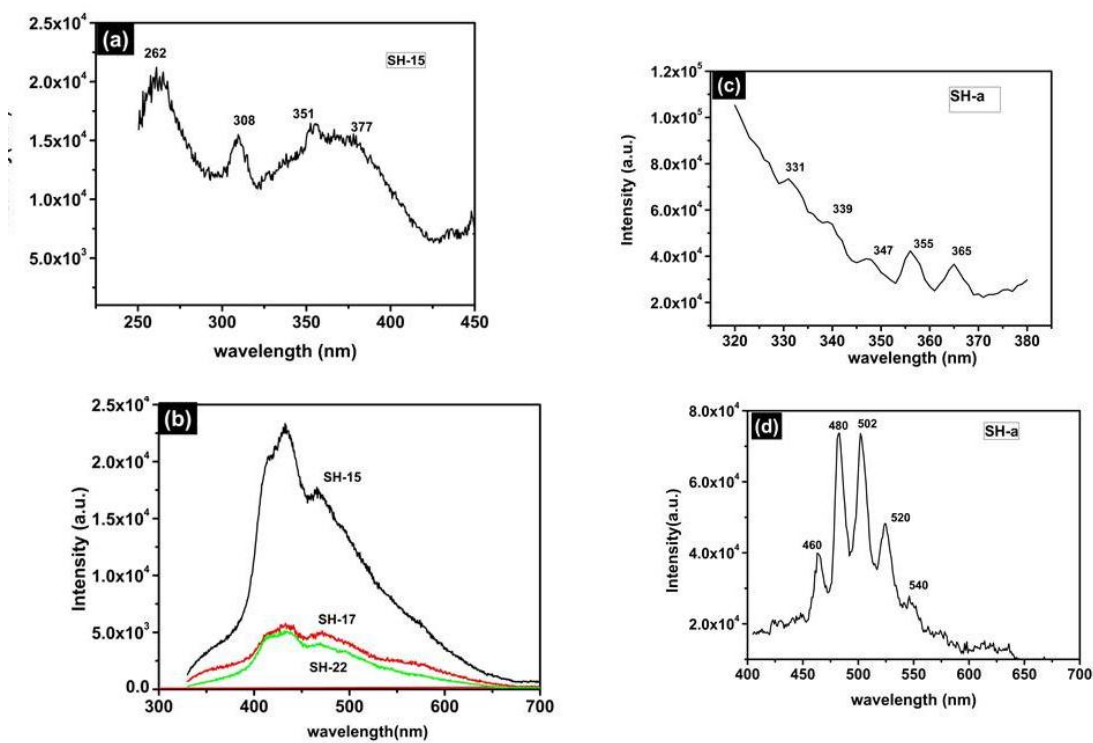

**Fig. 3** (a) PL excitation spectra of SH-15 (b) PL emission spectra of SH-15, SH-17 and SH-22 (c) PL excitation spectra of SH-a, (d) PL emission spectra of SH-a

The difference in the spectral luminescence of different uranium moieties helps in their detection and discrimination in complex rock matrix. Spectroscopic characteristic of uranium shows that it has absorption in the UV and near UV region and can emit in the violet blue region (390 - 490 nm) in the U(IV), and in the green region (490- 597 nm) in the U(VI) state [10,11,12]. PL studies were therefore utilized to identify specific uranium moiety in these ferruginous shale samples. Our earlier PL studies conducted on mineral concentrate sample of SH-9 have indicated the stabilization of both $U^{4+}$ and $U^{6+}$ in the FAP and haematite matrix. PL characterization in present study revealed the presence of $UO_6^{6-}$, $UO_2^{2+}$ and U(IV) species in different rock matrices.

***Whole rock samples SH-15, SH-17 and SH-22.***

The excitation spectrum of whole rock samples of SH-15, SH-17 and SH-22 sample is shown in Figure 3(a). The samples were excited by UV- vis light (245nm) corresponding to the basic and activated electronic and vibronic level of uranium and emission was monitored in the range from 250-700 nm. As per literature, uranium in the hexavalent state is known to exhibit excitation peaks at 326 nm (30674 $cm^{-1}$), 362 nm (27624 $cm^{-1}$), 392 nm (25510 $cm^{-1}$), 412 nm (24271 $cm^{-1}$), 424 nm (23584 $cm^{-1}$), 445 nm (22471 $cm^{-1}$), and 487 nm (20746 $cm^{-1}$) [13]. Hence the emission can be monitored using any of these peaks. Excitation peaks were observed at 262, 308, 351 and 377nm. The fluorescence spectra of SH-15, SH-17 and SH-22 shows unstructured luminescent blue green spectra without any vibronic progressions (Figure 3(b)). The maximum of the bands is near 435 nm and two shoulders are identified at 416 nm and 465 nm. The absence of vibronic structure clearly indicates that the transition intensity is from electronic origin due to substantial electric dipole change during electronic excitation at the low symmetry site indicating the fact that the luminescence is from the uranate group. Similar spectra have been observed by Mohapatra et al. [14, 15]. This implies the stabilization of uranium as octahedral $UO_6^{6-}$ group in the FAP matrix. The excitation spectra of the uranate ion is basically a charge transfer transition from orbital having mainly oxygen 2p character to orbitals having mainly uranium 5f orbitals while the emission spectra is a parity forbidden charge transfer transition 5f-$t_{1u}$ vibronically allowed by coupling with the ungerade vibration mode of the $UO_6^{6-}$ octahedron [16]. No correlation of the emission intensity with $U_3O_8$ content was observed. Since no secondary uranium mineral was identified in XRD hence the presence of U(VI) is attributed to the replacement of $Ca^{2+}$ ions in FAP.

The life time decay studies performed on SH-15 is shown in Fig 4(a). Upon 285 nm excitation the decay emission was observed at 434 nm corresponding to the peak emission observed in the fluorescence emission spectra. Similar emission spectra were also obtained for SH-17 and SH-22. The decay curves were fitted with a bi-exponential function using the equation $y = A_1 \exp(-t/\tau_1) + A_2 \exp(-t/\tau_2)$ where $A_1$ and $A_2$ are pre-exponential factors and $\tau_1$ and $\tau_2$ stands for the lifetime of fast and slow decays. The relative fraction for the fast and slow decays is indicated in brackets. The observed effective lifetime τ for bi-exponential decays for SH-15, SH-17 and SH-22 was observed to be $\tau_1$ = 1.05 μs (67%) and $\tau_2$= 11.25 μs (34%); $\tau_1$ = 1.11 μs (74%), $\tau_2$=9.82 μs (26 %); $\tau_1$ = 0.9 μs (77%), $\tau_2$=

10.4 µs (22%)  respectively. Since no other species was being detected in PL, hence the presence of biexponential decay confirms the stabilization of $UO_6^{6-}$ ion in two different coordination geometries. In the FAP matrix, due to the similar ionic radius of $U^{6+}$ (0.76 Å) with $Ca^{2+}$ (1.06 Å), $UO_6^{6-}$ ion can substitute either for Ca (I) in nine fold coordination the symmetry for which is $C_{3v}$. Next it can substitute for Ca(II) in sevenfold coordination in Cs symmetry. However $U^{4+}$ and $U^{6+}$ are most likely to be substituted in Ca (II) position [17]. This implies that both the sites are likely host for uranium ion. $UO_6^{6-}$ is a centrosymmetric group hence the electric dipole transition is otherwise forbidden due to higher symmetry. This leads to long emission times for uranates, of the order of 100's of microseconds in different host materials, however the decay time observed in our case is relatively shorter [18]. This can be attributed to both the occupancy of uranium in noncentrosymmetric site and charge compensation mechanism. The difference in the charges between the $(UO_6)^{6-}$ and $Ca^{2+}$ leads to multiple substitutions and vacancy creation at the subsequent sites for charge compensation. These structural defects act as quenching centers for luminescence which increases the nonradiative transition. Subsequently the PL radiative life time is reduced following the expression (1)

$$\frac{1}{\tau} = \frac{1}{\tau_R} + \frac{1}{\tau_{NR}} \quad \ldots\ldots\ldots\ldots (1)$$

Where $\tau_R$ and $\tau_{NR}$ are radiative and non radiative decay times.

*Luminiscent species SH-a*

Fig 3(d) shows the fluorescence spectrum of the sample SH-a recorded in the range of 430-700 nm upon 330 nm excitation. The emission spectra observed was similar to that of the uranyl $(UO_2)^{2+}$ group. The highly structured component appearing in the photoluminiscent spectrum of SH-a showed resemblance with the uranium U(VI) carbonate phase and uranium U(VI) phosphate phase. In oxidising conditions prevailing in phosphatic rich region, autunite is the widely prevalent principal mineral. Other minerals ocuuring in phosphatic and carbonate rich uraniferous regions are saleeite, leigbite, phosphouranylite. However the spectral features i.e. position of the zero phonon line, vibronic line spacing and vibronic intensity ratio do not match with any of the uranium phosphatic or carbonate mineral of the uranium spectral library. The elemental analysis of the species SH-b with EDS revealed Si, O and U. The presence of uranyl dioxo cation O=U=O, $(UO_2)^{2+}$ species determined by the luminiscent spectrum is similar to that observed previously . Several other studies have also reported similar luminescence of uranyl. For uranyl group these spectral effects are assigned to a Ligand to metal charge transfer (LMCT) transition where the electron transfer takes place from bonding oxygen orbital to non bonding uranium 5f orbital. Emission peaks corresponding to the vibronic bands were observed at 480 nm, 502 nm, 520 nm, 540 nm and 570 nm. The intensity and spacing of these vibronic bands is a function of the uranium ligand interaction. The zero phonon line corresponding to the first vibrational bond was observed at 502 nm in the emission spectrum. The stretching frequency corresponding to these vibronic band of O=U=O is observed at 862 cm$^{-1}$ which could be attributed to the coupling of the Raman active bond with the 3Πu electronic triplet excited state. The excitation spectra of the luminescent species SH-a revealed peaks at 331 nm, 339 nm, 347, 355 and 365 nm (Fig 3(c)). XRD and Raman spectroscopy has revealed that $(UO_2)^{2+}$ ion is adsorbed on silica or chalcopyrite .

Fig 4(b) represents the life time decay of SH-a at an excitation wavelength of 266 nm while monitoring the emission at 365 nm. The short and long life time decays corresponding to 44 μs and 110 μs with relative % of 69 and 31 are observed. The emission for $(UO_2)^{2+}$ is an electric dipole parity forbidden transition and decay times observed for this noncentrosymmetric group is of the order of 100's of μs, similar to that observed by other studies.

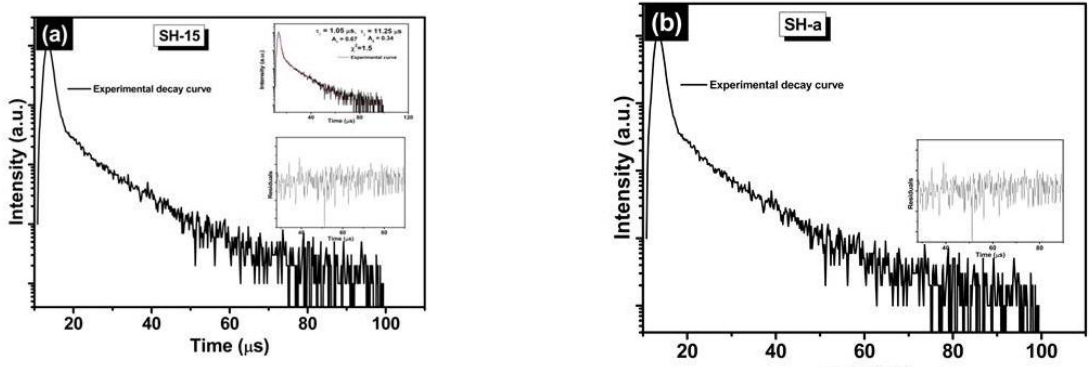

**Fig. 4** TRPL curve of (a) SH-15 (b) SH-a

Table 1 Physical charcateristics of samples.

| Sr. No | Sample | Colour |
|---|---|---|
| 01 | SH-9 | Dark Brown |
| 02 | SH-15 | Light Brown |
| 03 | SH-17 | Light red |
| 04 | SH-22 | Reddish brown |
| 05 | SH-a | Yellow Green |
| 06 | SH-b | Black |

## Conclusions

To summarise, by virtue of PL investigations the oxidation state and nature of uraniferous species of Bijawar region were effectively distinguished. Uranium was found to be present predominantly in the U(VI) oxidation state as uranate $(UO_6)^{6-}$ ion as a substitute for $Ca^{2+}$ in FAP however the signature of U(VI) as $UO_2^{2+}$ ions adsorbed on silica or chalcopyrite and U(IV) signature due to $UO_2$ were also found to be present. The fluorescence decay time was measured to be in the range of nanoseconds to microseconds. While the present of uranium in hexavalent oxidation state reveal a primarily oxidizing atmosphere, the presence of U(IV) oxidation state reveal the possibilty of weathering of initial uraninite present in this region. This study aims to provide a spectral database of uranium species in Bijawar region.

## Acknowledgements

Two of the authors are exteremely grateful to Shri A. K. Rai, Director AMD for granting permission to pursue this project.